\documentclass[pre,a4paper,onecolumn,notitlepage,nofootinbib]{revtex4-1}
\usepackage{amsmath}
\usepackage{bbm}
\usepackage{bm}
\usepackage{latexsym}
\usepackage{amsfonts}

\usepackage[pdftex]{graphicx}
\usepackage{subfigure}
\usepackage{epstopdf}
\usepackage{color} 
\usepackage[utf8]{inputenc}
\usepackage[T1]{fontenc}
\usepackage{times}

\newcommand{\beq}{\begin{equation}}
\newcommand{\eeq}{\end{equation}}

\newcommand{\<}{\left<}
\renewcommand{\>}{\right>}

\begin{document}

\title{\bf On the pros and cons of using temporal derivatives to assess brain functional connectivity}

\author{Jeremi K. Ochab}\email{jeremi.ochab@uj.edu.pl}
\author{Wojciech Tarnowski}\email{wojciech.tarnowski@uj.edu.pl}
\author{Maciej A. Nowak}\email{maciej.a.nowak@uj.edu.pl}
\author{Dante R. Chialvo}\email{dchialvo@unsam.edu.ar}

\affiliation{Marian Smoluchowski Institute of Physics and 
Mark Kac Complex Systems Research Center \\
Jagiellonian University, ul. \L{}ojasiewicza 11, PL30–348 Krak\'ow, Poland}
\affiliation{Center for Complex Systems \& Brain Sciences (CEMSC$^3$), Escuela de Ciencia y Tecnolog\'ia,  Universidad Nacional de San Mart\'{i}n, \& Consejo Nacional de Investigaciones Cient\'{i}ficas y Tecnol\'{o}gicas (CONICET),  25 de Mayo 1169, San Mart\'{i}n, (1650), Buenos Aires, Argentina}

\date{\today}

\begin{abstract}
The study of correlations between brain regions is an important chapter of the analysis of large-scale brain spatiotemporal dynamics. In particular, novel methods suited to extract dynamic changes in mutual correlations are needed. Here we scrutinize a recently reported metric dubbed ``Multiplication of Temporal Derivatives'' (MTD) which is based on the temporal derivative of each time series. The formal comparison of the MTD formula with the Pearson correlation of the derivatives reveals only minor differences, which we find negligible in practice. A comparison with the sliding window Pearson correlation of the raw time series in several stationary and non-stationary set-ups, including a realistic stationary network detection, reveals lower sensitivity of derivatives to low frequency drifts and to autocorrelations but also lower signal-to-noise ratio. It does not indicate any evident mathematical advantages of the proposed metric over commonly used correlation methods.
 
\medskip


\end{abstract}

\maketitle
 
\section{Introduction}
An important challenge and a very active area of research in the neuroimaging community is the study of correlations between brain regions, as a central point in the analysis of large-scale brain spatiotemporal dynamics. 
Usually the correlation measures are computed between several thousand time series---the  BOLD (``blood oxygenated level dependent'') signals---covering the entire brain, and averaged over a period of tens of minutes. However such evaluations fall short of characterizing the highly dynamic changes occurring in the brain\cite{Chang,preti,yu,Gonzalez,Thompson_Fransson} and, consequently,  the current emphasis shifted to the study of time-varying correlations.

 
In that line, a recent report~\cite{Shine2015} introduces a metric dedicated to investigation of dynamic functional connectivity (DFC) in fMRI time series data---dubbed Multiplication of Temporal Derivatives (MTD)---based on the temporal derivative of each BOLD time series.
The authors used this measure on a three-part experiment claiming that it demonstrates ``the ability of this novel metric to calculate dynamic and stationary functional connectivity structure in both real and simulated data''~\cite{Shine2015}.
Naturally, there are numerous other DFC methods available, based on sliding-window Pearson correlation (e.g., \cite{Chang,Hutchison}), clustering \cite{Liu} or temporal ICA \cite{Smith2012} to name just a few.
Many of them can be formulated in a general conceptual framework described by Thompson and Fransson \cite{Thompson2017a}---most notably MTD method and weighted Pearson correlation to which it bears resemblance.
It is of particular importance to explore and determine in which tasks these methods perform well and what properties of BOLD time series they rely on.

These notes are dedicated to carefully inspect the mathematical grounding of the MTD measure and revisit some of the scenarios for which the metric is intended.
We will first discuss the case of stationary processes, and then inspect the non-stationary cases.
The paper is organized as follows: the next section contains the mathematical definition of the MTD measure side by side with the mathematical expression for the Pearson correlation coefficient.
To develop some intuition, Section \ref{sec:examples} provides extremely simple examples allowing comparison of the expected results for Pearson correlation in the raw time series and its time derivative. Section \ref{sec:toymodel} presents a formal framework to predict the expected behavior of the correlations for the case in which a raw time series is compared with its temporal derivative. In Section \ref{sec:VII} we inspect a simple non-stationary example which lends support to the analytical expectations for the two time series (raw and derivatives).
The analytical framework is further contrasted, in Section \ref{sec:bold}, with the results of analyzing resting state BOLD time series. In Section \ref{sec:bold2} we use surrogate data from BOLD time series to determine the behavior of the derivatives to a sudden change in covariance. Finally, in Section \ref{sec:network} the performance of the MTD approach in inferring the underlying network connectivity is examined.  The paper closes with conclusions and a brief summary of the results. Further details are condensed in the Appendix section with the appropriate derivations.

\section{Correlations of first-order derivatives}
\label{sec:1}

Let us start by recalling how the authors in~\cite{Shine2015} define Multiplication of Temporal Derivatives:

\begin{align}
ds_{it}& = s_{it+1}-s_{it}\label{eq:1}\\
MTD_{ijt}& = \frac{ds_{it}ds_{jt}}{\bar{\sigma}_{i}\bar{\sigma}_{j}}\\
SMA_{ijt}& =\frac{1}{2w+1}\sum_{t'=t-w}^{t+w}MTD_{ijt'}=\frac{1}{2w+1}\sum_{t'=t-w}^{t+w}\frac{ds_{it'}}{\bar{\sigma}_{i}}\frac{ds_{jt'}}{\bar{\sigma}_{j}},\label{eq:shine}
\end{align}
where $2w+1$  equals to the number of samples considered in a temporal window $[t-w,t+w]$, $s_{i}$ is an $i$-th time series, and $\bar{\sigma}_{i}$ is the standard deviation of the entire $ds_{i}$ series. 
We call $s_{it}$ the \textit{raw} time series as opposed to the series of derivatives $ds_{it}$.
Usually, the series $ds_{it}$  is called interchangeably increments, finite differences, temporal derivatives or differentiated time series.
While the first two are terminologically more adequate,
for consistency with the established name of MTD we continue to use \textit{temporal derivatives} throughout.
Note that \eqref{eq:1} is a forward difference with a unit time step, but other choices are also possible.

Now, let us reflect on how these definitions relate to correlation coefficients. For any two time series $s_{it'}$ and $s_{jt'}$,
where $i,j \in \{1,2,\ldots, N\}$ are indices numbering the series and $t' \in \{t-w, t-w+1,\ldots, t+w-1,t+w\}$ is a time index in a window around time $t$,
the sample estimator of Pearson correlation coefficient for the given time window is defined as
\beq
\label{eq:Rdef}
r_{ijt}=\frac{\sum_{t'=t-w}^{t+w}(ds_{it'}-\bar{ds}_i)(ds_{jt'}-\bar{ds}_j)}{\sqrt{\sum_{t'=t-w}^{t+w}(ds_{it'}-\bar{ds}_i)^2}\sqrt{\sum_{t'=t-w}^{t+w}(ds_{jt'}-\bar{ds}_j)^2}}=\frac{1}{2w+1}\sum_{t'=t-w}^{t+w} \frac{ds_{it'}-\bar{ds}_i}{\sigma_i}\frac{ds_{jt'}-\bar{ds}_j}{\sigma_j},
\eeq
where $\bar{ds}_i=\frac{1}{2w+1}\sum_{t'=t-w}^{t+w}ds_{it'}$ is the sample mean and $\sigma_i$ denote standard deviations of the time series $ds_{i}$ in the given time window $[t-w,t+w]$.
Centering around the mean and dividing by standard deviations is necessary for the coefficient to be invariant under  linear transformations ($a+b ds_{it}$, with constants $a$ and $b$).

\begin{figure}[b!]
\centering
\includegraphics[width=0.65\textwidth]{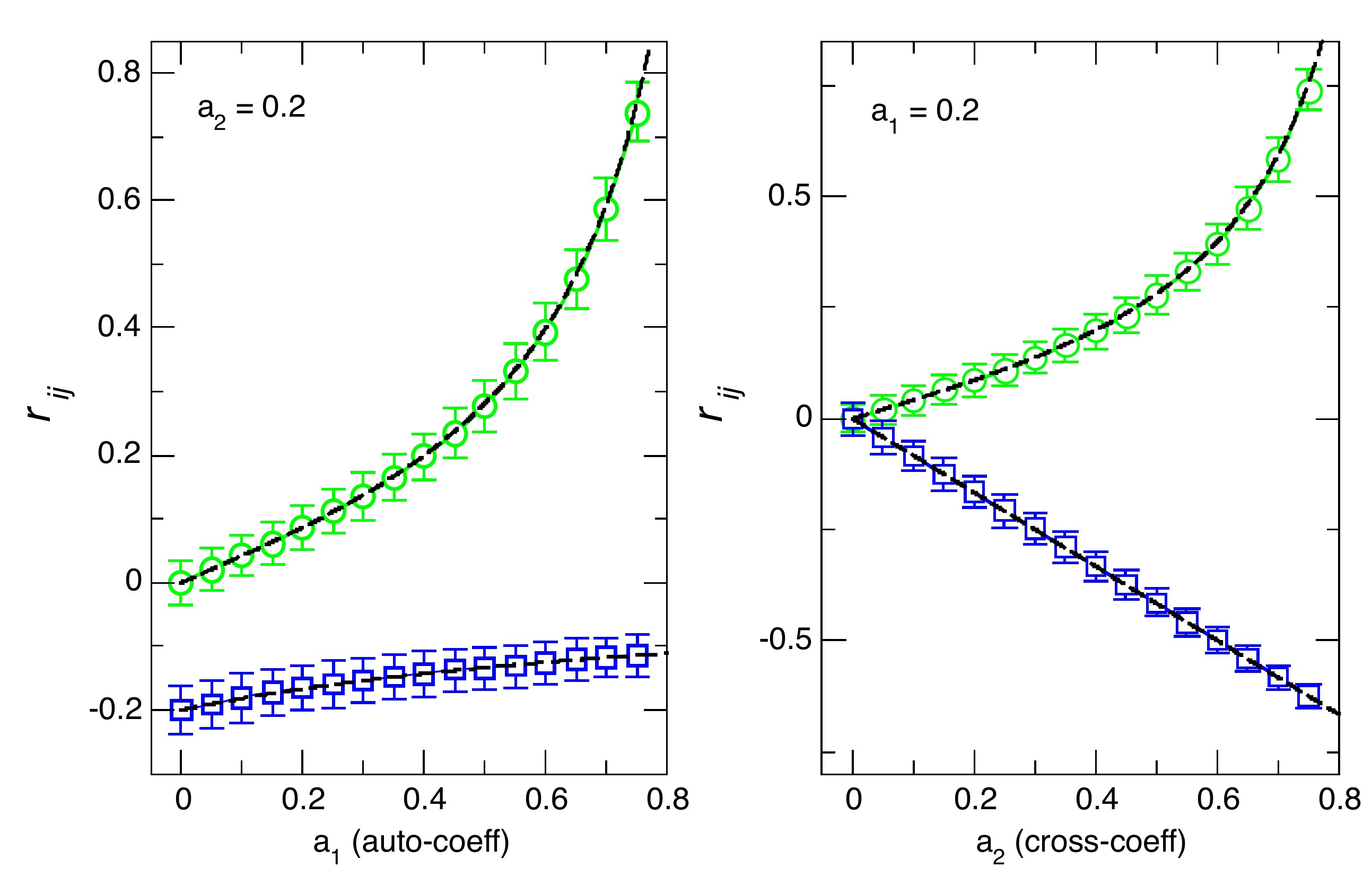}
\vfill
\includegraphics[width=0.65\textwidth]{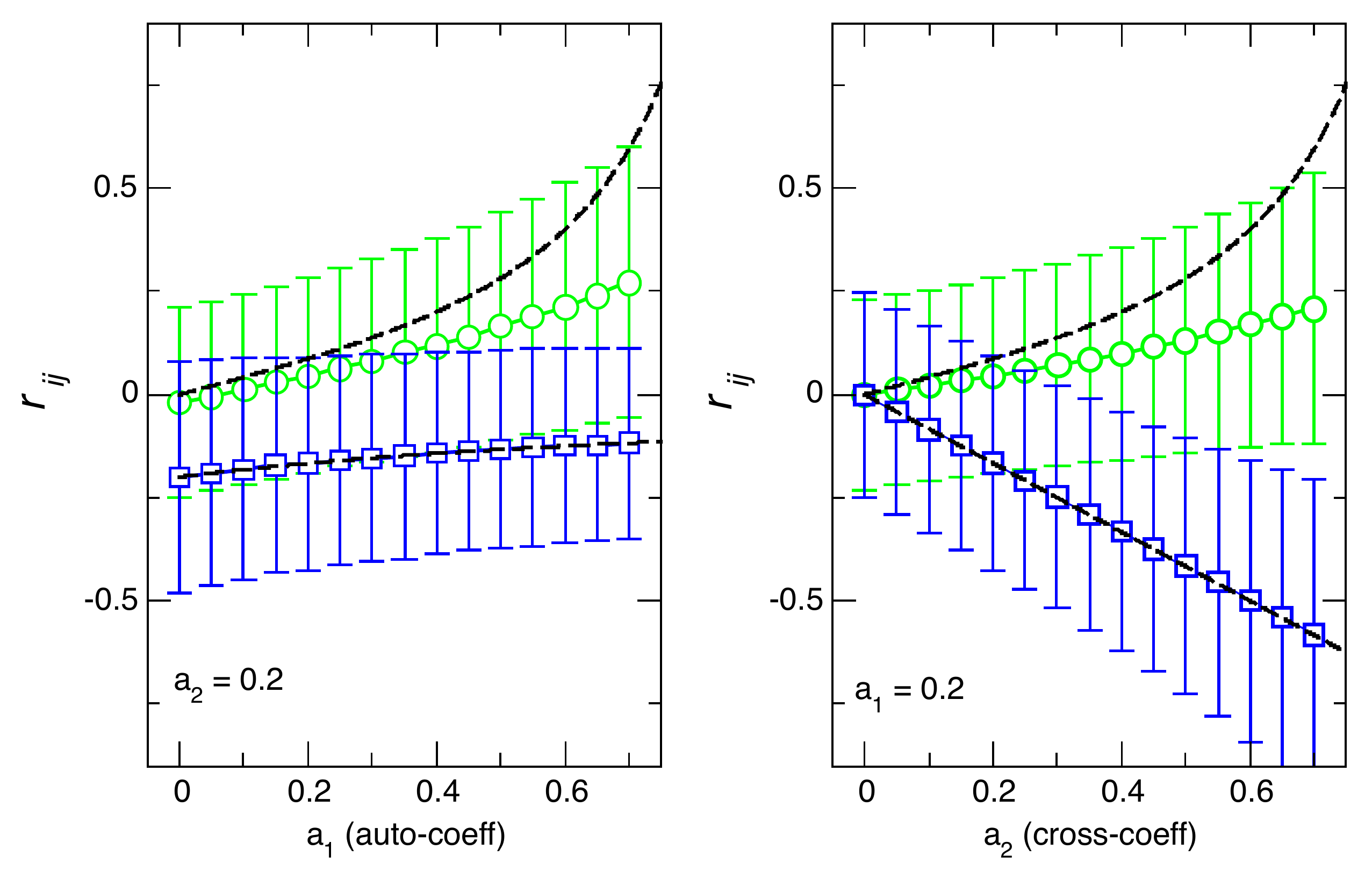}
\caption{\label{fig:2} Analytical and numerical results for the correlations of a AR(1) process of two nodes with a range of coefficients (the auto-coeff. $a_1$, and the cross-coeff. $a_2$). In all graphs $r_{ij}$  corresponds to the Pearson correlation for the raw (green circles) and derivative (blue squares) time series. Dashed lines represent the asymptotic analytical expectation for $r_{ij}$, and symbols and error bars correspond to the mean and standard deviations of the numerical results.
Upper panels show correlation of entire series, lower panels show sliding window  correlation (i.e., SWPC of raw series and MTD).
Time series length $T=1000$; $N=1000$ realizations; window length $2w+1 = 21$.
}
\end{figure}

The form of the equations \eqref{eq:shine} and \eqref{eq:Rdef} are visibly similar. The difference is that the derivatives in $SMA_{ijt}$ are not centered and that the standard deviations $\bar{\sigma}_i$ in \eqref{eq:shine} are computed over the whole series and not just over the time window as $\sigma_i$ in \eqref{eq:Rdef}. As regards centering, the mean $\bar{ds}_i=\frac{1}{2w+1}\sum_{t'=t-w}^{t+w}ds_{it'}=\frac{s_{it-w}-s_{it+w}}{2w+1} \rightarrow 0$ for large window sizes; the standard deviation of the (window) sample also converges to the standard deviation of the entire series.
While we expect that for short windows the centering and variances might introduce some bias,
Figure \ref{fig:2} corroborates that the two ways are tantamount: the numerical results for temporal derivatives calculated exactly from \eqref{eq:shine} conform to the analytical solution derived for large window limit of \eqref{eq:Rdef}.

In other words, asymptotically for large window sizes the SMA of MTD~\cite{Shine2015}, i.e., the moving average of multiplication of temporal derivatives equates with the sliding-window Pearson correlation (SWPC) of temporal derivatives (not to be confused with the SWPC of the raw series).
In the general notation used by Thompson and Fransson \cite{Thompson2017a}, $Y=R(U(X);W)$---where $X$ are raw data series, $W$ are weight vectors, $U$ is a transform of the data, $R$ is a relation function, and $Y$ is the resulting estimate of DFC---SMA of MTD can be expressed with the temporal derivative \eqref{eq:1} standing for $U$ and Pearson correlation standing for $R$. While we believe, the ``SWPC of derivatives'' is a more informative term than MTD, hereafter we do not use it in order to avoid terminological confusion.

As also mentioned in \cite{Shine2015}, differencing has the high-pass filtering effect preferable in some situations. 
Indeed, differencing is a well-known method for reducing some time series to stationary ones \cite{BoxJenk}, which one could expect to show no or, in real data, at least less dynamics.
Therefore, a valid question is how is this method different from the sliding window Pearson's correlation coefficients referred to therein or, from another angle, what information do derivatives contain that the raw series do not?
To unravel this issue, in the next section we will first scrutinize the simplest examples of time series, allowing a comparison with the Pearson correlation expected for a raw time series and its temporal derivatives. Again we remark that the stationary cases will be treated first to later analyze the relevant case of non-stationarity. 
\begin{figure}[b!]
\centering
\includegraphics[width=0.55\textwidth]{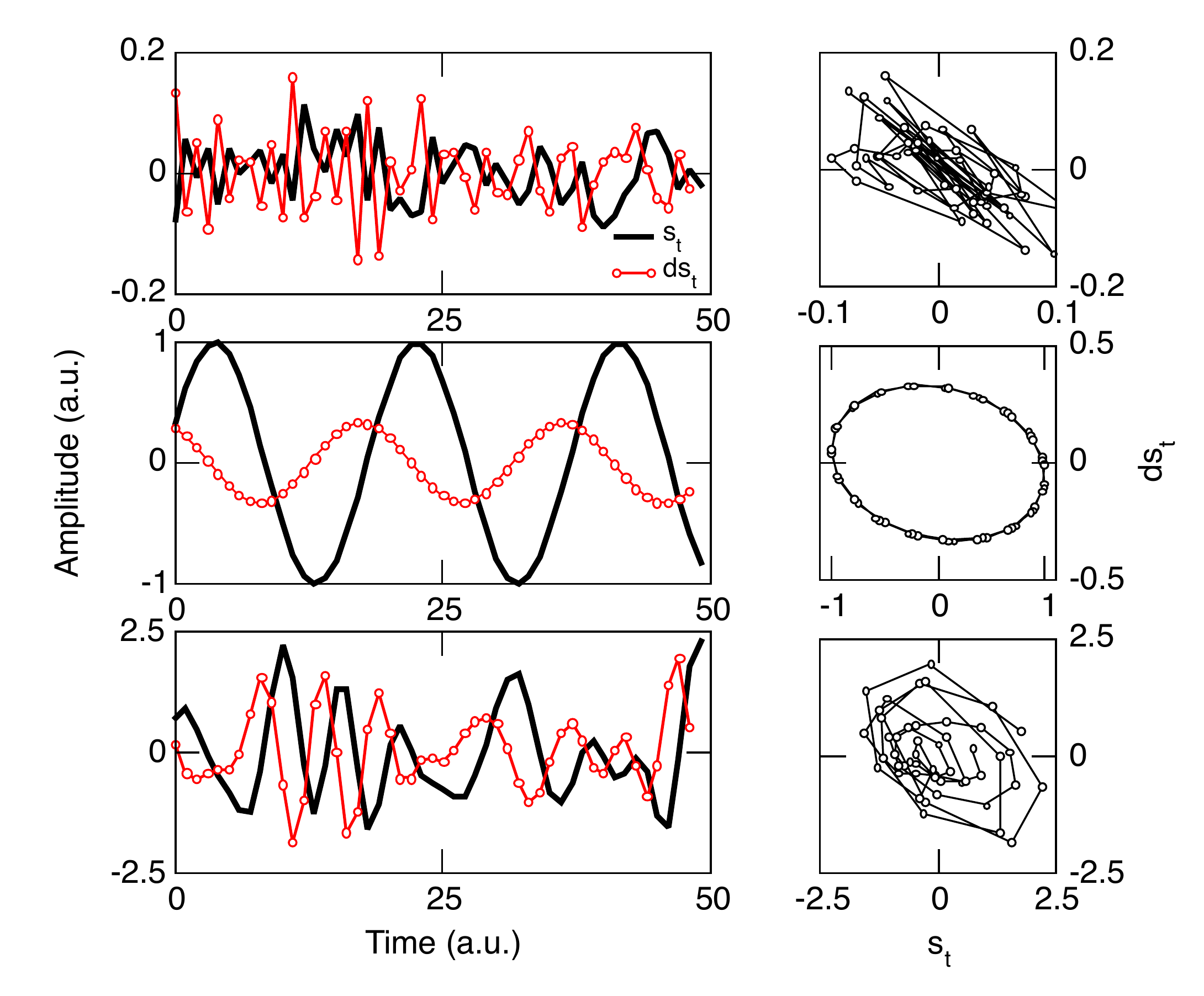}
\caption{\label{fig:1} Simple examples of raw time series $s_t$ (black) and its series of derivatives $ds_t$ (red) for a Gaussian (top), a sinusoid (middle), and a typical brain BOLD (bottom) time series. The diagrams on the right plot the time series against each other.}
\end{figure}

\section{Simple examples}
\label{sec:examples}
For the sake of simplicity, let us consider how the two correlation measures behave in simple cases.
First, in the case of signals which are Gaussian white noise (see Fig.~\ref{fig:1}) let us assume that $s_{it} = \xi_{it}$
where $\xi_{i}$ are independent and identically distributed random variables, and the $\xi_{it}$ are the outcomes of these variables (random variates).
Let us remind that independence of random variables $X, Y$ is equivalent to the statement that the expectation value factorizes $E[f(X)g(Y)]=E[f(X)]E[g(Y)]$ for any functions $f$ and $g$. 
In the case of Pearson correlation, the centered time series $E[s_{it}-\bar{s}_{i}]=0$ and thus, thanks to the independence between $s_{i}$ and $s_{j}$, also the correlation coefficient \eqref{eq:Rdef} is zero for $i \neq j$. 
It follows that since $E[ds_{i}]=0$ also $E[ds_{it}ds_{jt}]=0$ for $i\neq j$, and so the expected value of the moving average \eqref{eq:shine} yields $E[SMA_{ijt}]=0$ as well.
It is noteworthy, however, that since $ds_{i}$ is a sum of two random variables, its variance is twice the variance of $s_{i}$. Thus, in all noisy time series we can expect the signal-to-noise ratio of the derivatives to be lower than that of the raw series.

Now let us assume that instead of a discrete time series $s_{it}$, we have a continuous, differentiable signal $s_{i}(t)=\sin(t + \phi_i)$, which has zero mean and $\sigma_i^2=1/2$.
For signals $s_{i}(t),s_{j}(t)$ differing only by their phases $\phi_i, \phi_j$ the correlation coefficient equals $r_{ij}=\cos(\phi_i-\phi_j)$. The differentiated series $ds_{i}(t)=\cos(t + \phi_i)$ have the same mean and variance and correlation coefficients. Introducing arbitrary frequencies have no effect as well. Thus, we expect similar results for the both Pearson correlation measures for signals that are sinusoidal.
For a discrete signal, the amplitude of derivatives decreases linearly with increasing sampling rate, see Fig.~\ref{fig:1} (middle panel).
Computationally, differences between the two correlation measures will only appear from an artifact when the length of the sliding windows is comparable to the period of the sinusoidal oscillation. In such case, there is an asymmetry, as the sine waves will be cut at different positions depending on their phase and frequency.
For the sake of intuition, the bottom panel of Fig.~\ref{fig:1} depicts a BOLD time series and its derivative, which illustrate the phase shift mentioned above for the case of sinusoidal signals.
Note that the derivatives produce series one data point shorter than the raw time series, as visible in the figure.

Before proceeding further let us recall that the main claim of~\cite{Shine2015} was that the cross-correlation of the derivative time series is more informative of brief correlation changes. 

\begin{figure}[t!]
\centering
\includegraphics[width=0.42\textwidth]{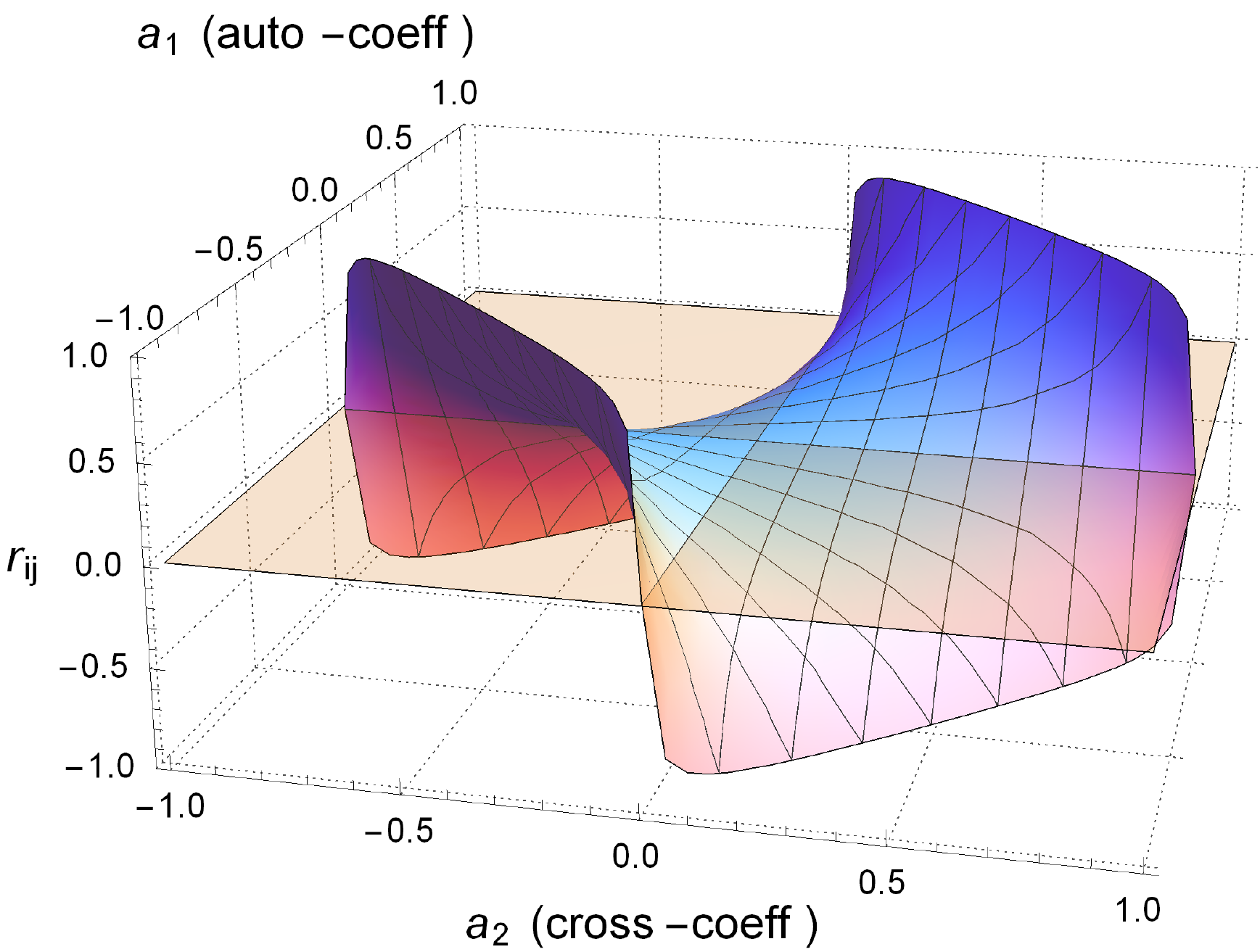}
\includegraphics[width=0.42\textwidth]{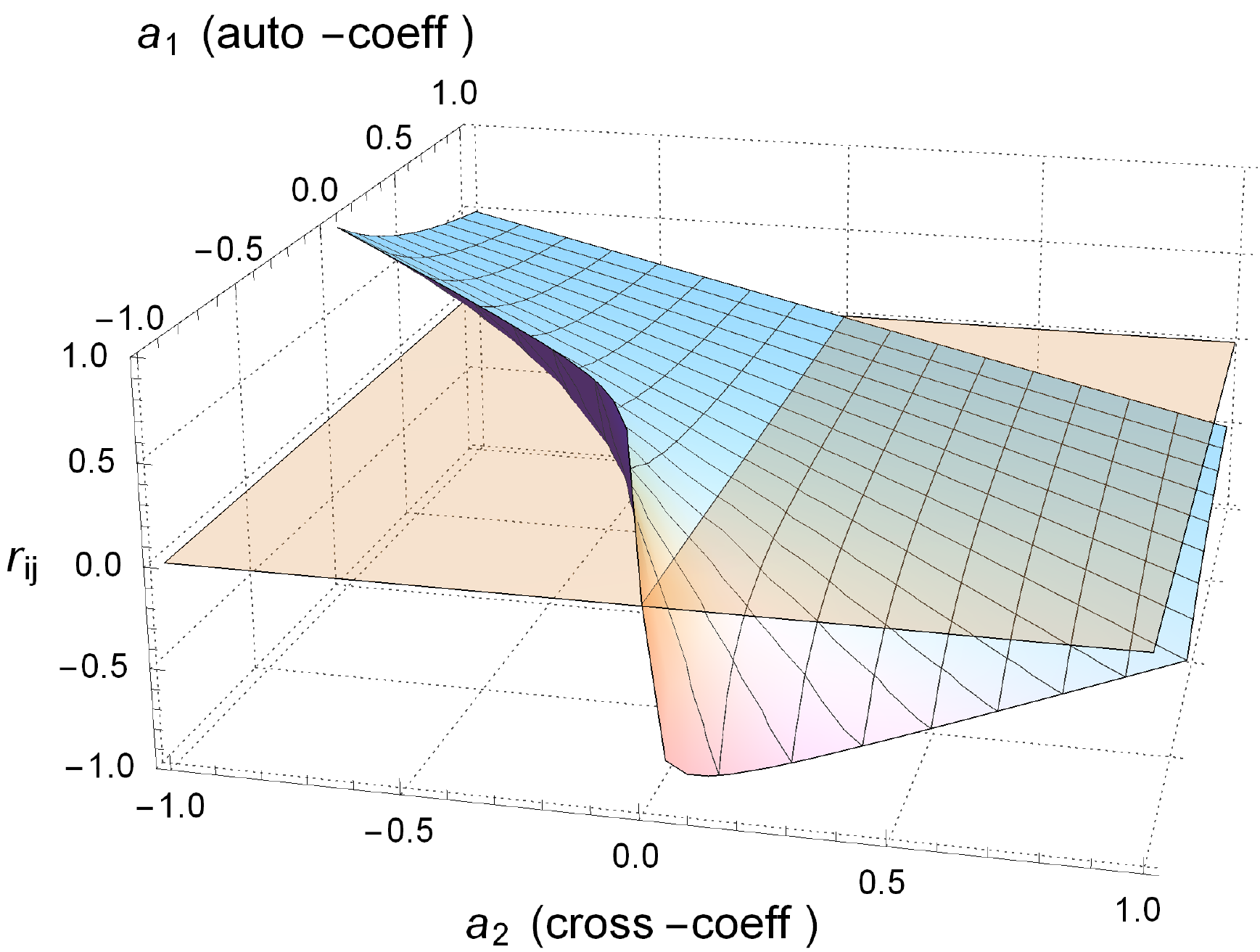}
\caption{\label{fig:3} Analytical asymptotic Pearson correlations of an AR(1) process of two nodes with a range of coefficients, as in Fig.~\ref{fig:2}. Left panel corresponds to raw series, right panel to temporal derivative. Note that the allowed parameter range is given by $|a_1|+ |a_2|< 1$.}
\end{figure}

\section{Autoregressive models}
\label{sec:toymodel}

To revisit and understand the results under scrutiny, in this section we study the case of two AR(1) processes for which we can manipulate their interactions and compute the expected correlations measures for both the raw and derivative time series.
Despite the enormous differences between the properties of the BOLD signal and a simple AR(1) process, its analysis can bring about some general understanding on what to expect from the proposed correlation metric of derivatives. The two time series are simulated as follows:
\begin{align}
\begin{split}
\label{eq:2series}
x_{1t}&=a_1 x_{1,t-1}+ a_2 x_{2t-1} + \xi_{1t}, \\
x_{2t}&=a_1 x_{2,t-1}+ a_2 x_{1t-1} + \xi_{2t}, 
\end{split}
\end{align}
where $\xi_{it}$ are uncorrelated.
The range of parameters is limited by $|a_1|+ |a_2|< 1$ to ensure weak stationarity.
The model \eqref{eq:2series} is a special case of VAR(q) processes---in the Appendices~\ref{app:1}-\ref{sec:vectorisation} we show how one can derive analytically and compute Pearson correlation (and correlation of the derivative) of such a process knowing its parameters.
Thanks to that, we can predict the average behavior of SWPC and MTD for a range of parameters within that model, as well as we can reverse-engineer the real data, designing a model that exhibits specific Pearson correlations. 

Notice that $a_1$ and $a_2$ above represent the auto-\emph{interaction} and the cross-\emph{interaction} coefficients, respectively, which can be estimated in this \emph{linear} case by the \emph{correlation} coefficients. Thus, in the jargon of~\cite{Shine2015} $a_1$ and $a_2$ are the values representing the ``ground truth'' which the numerical methods of functional connectivity shall predict.

Figure \ref{fig:2} (upper panels) shows the numerical results (symbols and error bars) for the asymptotic correlations of the process of two nodes in \eqref{eq:2series} as well as the analytical results (dashed lines) provided in Appendices~\ref{app:1}-\ref{sec:vectorisation}. The results show that in the case of fluctuations modeled by an AR(1) process the Pearson correlation of its derivative exhibits a negative sign. Moreover, depending on $a_1$ and $a_2$ it can be weaker or stronger than the correlation of its raw time series. A realistic, with high values of autocorrelation, is further shown in Sections~\ref{sec:bold}-\ref{sec:bold2}.
The analytical result for the full parameter range is shown in Fig.~\ref{fig:3}.

Similar conclusions can be drawn from computing sliding windows of the Pearson correlation of both the raw time series and its derivative as shown in lower panels of Figure~\ref{fig:2}. The only difference with the results in upper panels are the larger magnitude of the error bars expected from the relatively smaller sample size. 

\begin{figure}[t]
\centering
\includegraphics[width=0.45\textwidth]{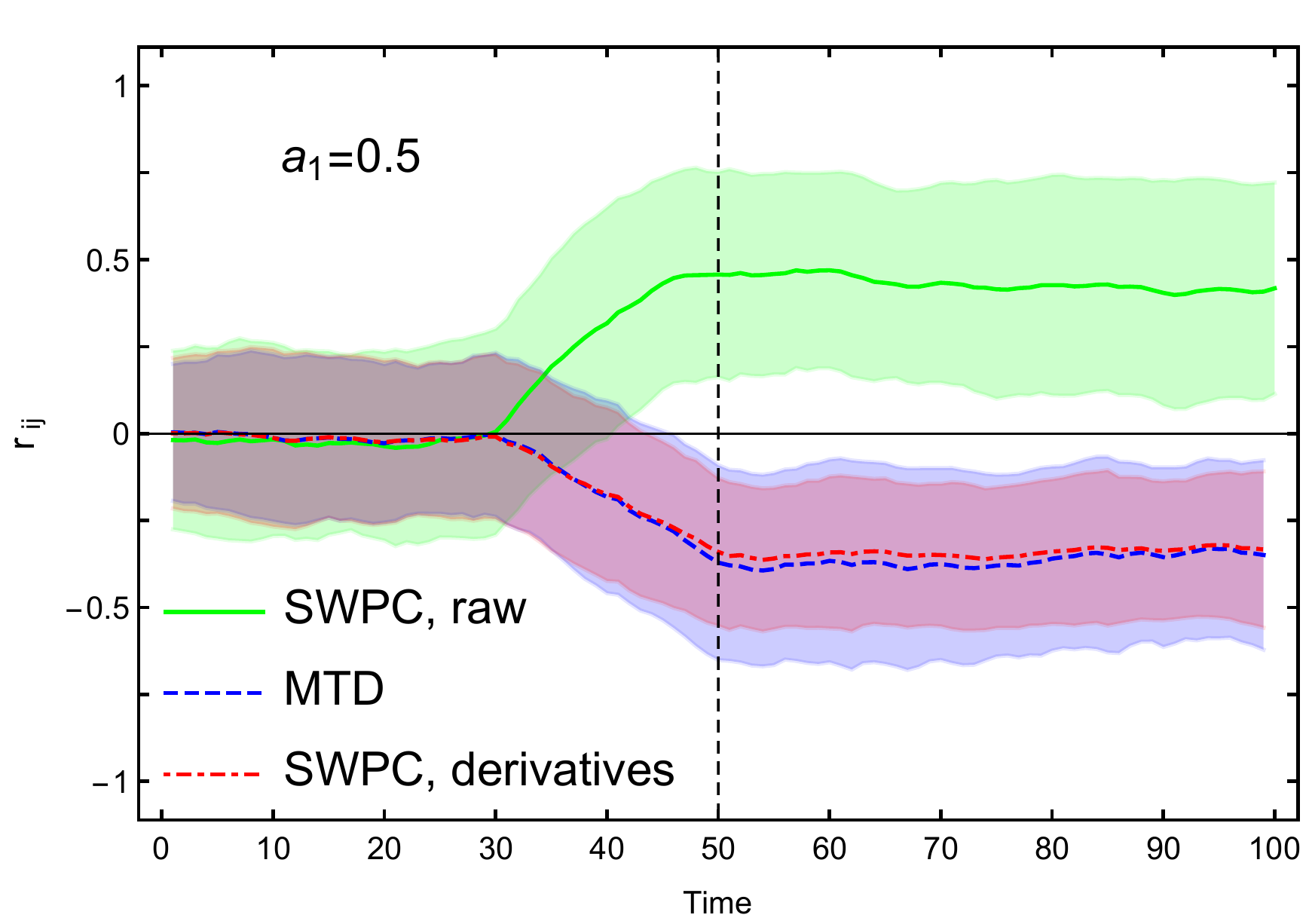}
\hfill
\includegraphics[width=0.45\textwidth]{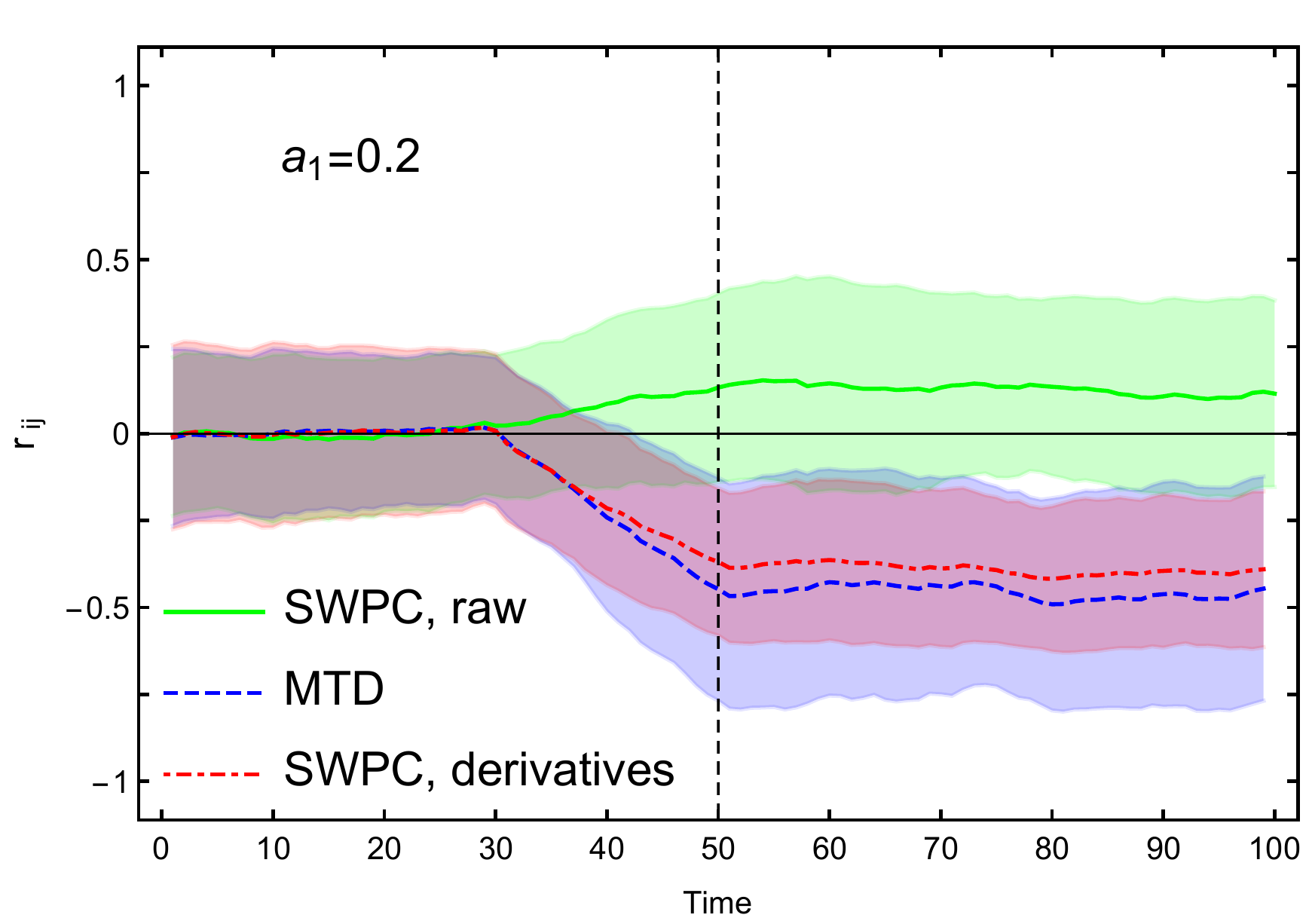}
\caption{\label{fig:6}A step change in  the cross-interaction coefficient  ($a_2(t)=0$ for $t<50$ and $a_2(t)=.5$ otherwise) can be detected by evaluating either the correlation of the raw or the derivative time series. Compare with Fig.~\ref{fig:3} to see the expected change in $r_{ij}$ for different values of auto-coefficient $a_1$. (Window size=21, sliding with unit steps, lines are averages and shaded regions are standard deviations of N=200 realizations).}
\end{figure}
 
\subsection{A non-stationary example}
\label{sec:VII}

The report introducing MTD~\cite{Shine2015} relies on the hypothesis that the time derivative of a signal could bring about new information in the case of time dependent mutual correlations. In simple terms, the idea was that sudden changes in correlations would be best estimated by the correlation of derivatives. The set-up of Experiments 1a and 1b therein is essentially Gaussian signals undergoing an instantaneous change from null to positive Pearson correlation (from 0.1 up to 0.5).
Autoregressive models offer a more general and slightly more realistic scenario, allowing us to tune the autocorrelations of the signal.
We consider the same scheme of \eqref{eq:2series}, with $a_1=0.2$ and $a_1=0.5$ but with the modification that the cross coefficient $a_2(t)$ depends on time, and is undergoing a sudden step change:
\begin{align}
\begin{split}
x_{1t}&=a_1 x_{1,t-1}+ a_{2}(t) x_{2t-1} + \xi_{1t}, \\
x_{2t}&=a_1 x_{2,t-1}+ a_{2}(t) x_{1t-1} + \xi_{2t}.
\end{split}
\end{align}

Then we compute correlation measures over the raw and derivative time series and attempt to predict at which time step the change occurs.
Figure \ref{fig:6} shows an example of the typical results obtained.
Mark that for the derivatives, we compare MTD \eqref{eq:shine} and SWPC \eqref{eq:Rdef}, showing that despite the difference in centering and standardization they behave almost the same even for a small window size.

It is visible that correlations computed over the raw or the derivatives accurately detect the change in the coefficient from $a_2=0$ to $a_2=0.5$ as soon as the sliding window reaches the transition point.
In fact, the correlation values after the transition can be read already from Fig.~\ref{fig:3} by looking at the corresponding parameter coordinates.
Moreover, from the preceding section, and specifically from Figures~\ref{fig:2}-\ref{fig:3}, one can see that the size of the change in correlations $r_{ij}$ heavily depends on the parameters $a_1$, $a_2$.
Also the relative size of $r_{ij}$ between raw series and derivatives depends on them.
Additionally, mark the standard deviations in Fig.~\ref{fig:6}: for $a_2=0$ they are comparable between the methods; for $a_2=0.5$, MTD has the largest spread, followed by SWPC of raw series and SWPC of derivatives having the best precision.
Lastly, while the raw series does not allow significant detection of change in cross-coefficient for small auto-coefficient, for high $a_1$ the detection is possible even sooner than for derivatives, owing to the steeper early slope during the transition.

\begin{figure}[b]
\centering
\includegraphics[width=0.8\textwidth]{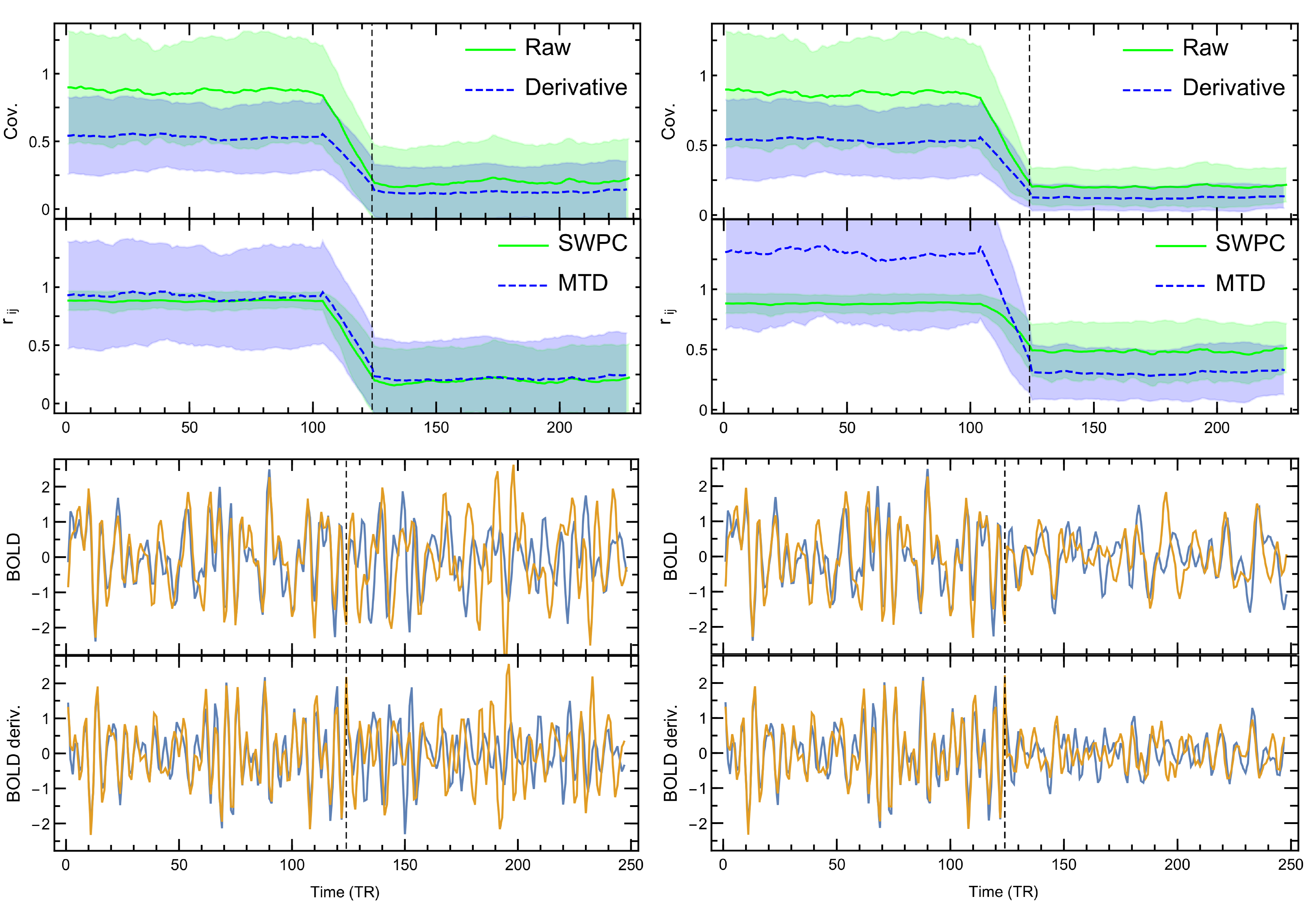}
\caption{\label{fig:7}Simulation of a sudden change (at time=124) in covariance using surrogate times series. Left panels: cross-correlation drops to 0.2 with constant autocorrelation close to 1. Right panels: cross-correlation drops to 0.2 and autocorrelation to 0.4. The bottom graphs depict the time series and derivatives of two selected ROI. The top graphs show the covariance and correlation as in Fig.~\ref{fig:6}, with sliding window of 21 steps. Lines are averages and shaded regions are standard deviations of N=200 realizations. Dashed vertical line indicates the step at which the change occurred.}
\end{figure}

This simple example demonstrates how strongly correlation metric of both raw series and derivatives depends on the model of the signal, which raises doubts about any general advantages of one method over the other to detect such non-stationary effects.
In some scenarios, however, one or the other method might have an edge---like MTD in the particular model of head motion originally reported in~\cite{Shine2015}.
The following sections, and in particular Fig.~\ref{fig:7}, test these observations in a much more realistic, data-driven fashion.

\begin{figure}[b!]
\centering
\includegraphics[width=0.65\textwidth]{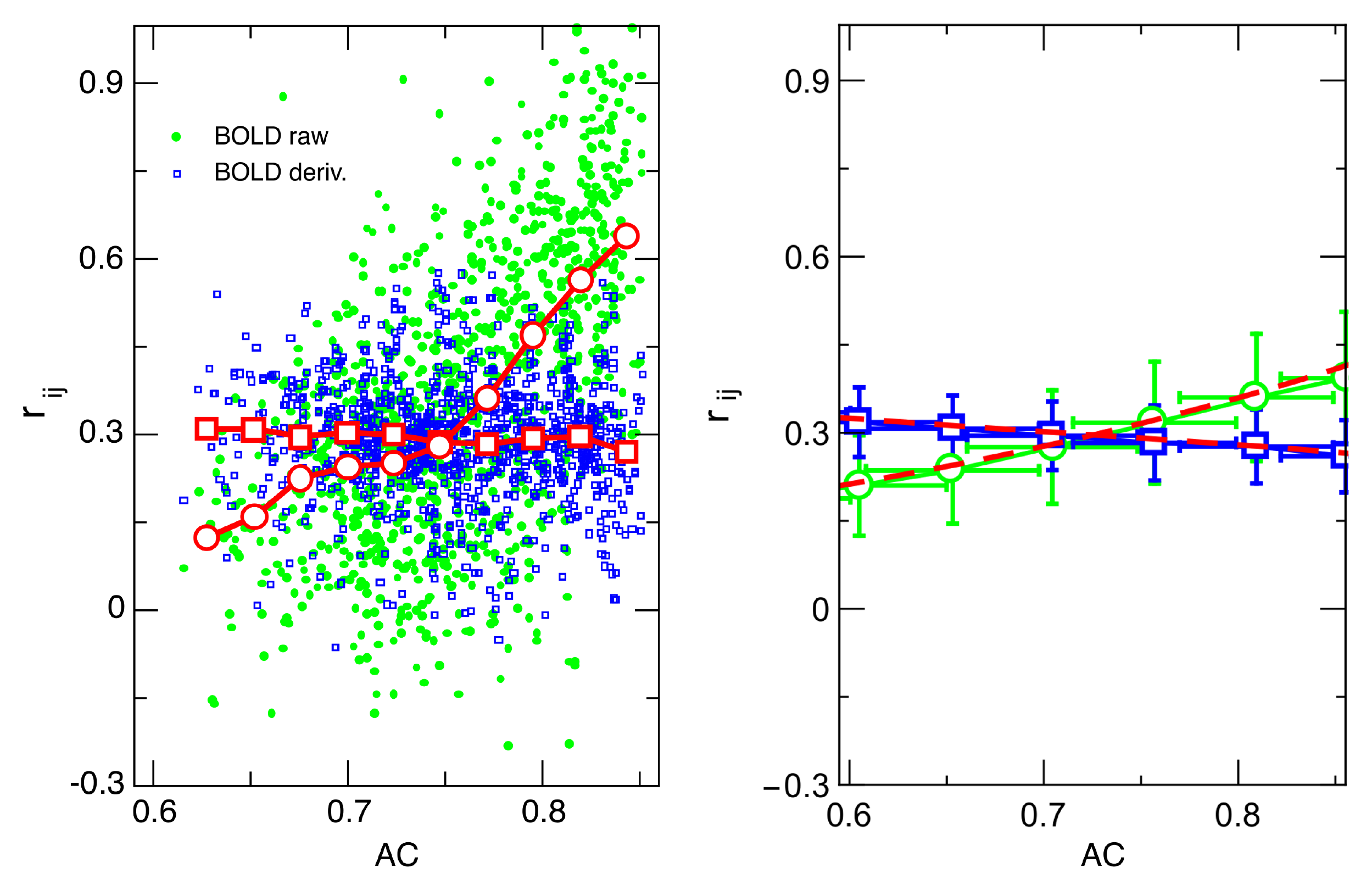}
\caption{\label{fig:4} Comparison between an ARMA model correlation behavior for both raw and differentiated experimental BOLD time series. Left Panel: Each dot represents the cross-correlation  ($r_{ij}$) of  two BOLD time series versus the average autocorrelation value (AC) of the pair (small filled green circles for raw BOLD and open squares for derivatives; 
red big circles denote binned averages for raw and squares for derivatives, with horizontally equidistant binning).
Right Panel: Dashed lines are analytical expectations of cross-correlations for the fitted model given its autocorrelation.
The symbols with errors bars correspond to mean cross-correlation values and standard deviations, respectively, from simulation of the fitted ARMA model, where circles denote results for the raw series and squares those from differentiated series.
The model consists of two coupled ARMA(1,1) time series with three $2\times 2$ coefficient matrices in total ($A$ for AR, $B$ for MA and $\Sigma$ for noise). The parameters $A_{1,1}=A_{2,2}=a_1$ were fixed by the autocorrelations of the BOLD pairs (green circles data points); the other parameters were fitted by least square differences between the analytical lines and the data points (both raw and differentiated).}
\end{figure}

\section{Correlation properties of BOLD signals and its fitted ARMA model}
\label{sec:bold}

Now we turn to study the correlation properties of real BOLD brain signals estimated for both the raw time series and its derivative. Furthermore, we extend the analysis of each time series to its autoregressive moving average (ARMA) models~\cite{arma}, an approach which is often used to explore the statistical properties of time series. This approach allows for a description in terms of a stochastic process with two polynomials, one for the auto-regression and the second for the moving average.
 
In Figure \ref{fig:4} we show the results of such an approach informed by brain resting state BOLD data. These time series correspond to closed-eyes resting state fMRI, 240 volumes with 2.5 s TR amounting to 10-minute recording of one healthy subject, already described in~\cite{Haimovici} (all the acquisition and preprocessing information can be found in the supplemental information therein) and were extracted according to the parcellation of~\cite{hagman} which covers the entire human cortex in $N=998$ patches of approx. $1\times 1\,$centimeters. In order to make a comparison of auto and cross-correlation properties, the autocorrelation of each time series was computed and used to sort the 998 data sets. After that the cross-correlations were computed between consecutive pairs of time series with neighboring autocorrelation values. The dots in Figure \ref{fig:4} represent the cross-correlation $r_{ij}$ between each pair as a function of their mean autocorrelation value. Green dots correspond to the raw data and blue dots to the time series of derivatives, while the symbols and error bars denote the predictions of the ARMA model.

The reason for scrutinizing the dependence of cross-correlation on autocorrelation is the following: in general, one would like to know, whether (linear) Pearson correlations of raw or derivative time series are informative of interactions between brain regions that produced the series.
In Sec.~\ref{sec:toymodel} we have shown that it is not only the interactions (modelled by cross-coefficient $a_2$) but also the auto-coefficient $a_1$ responsible for autocorrelations that might strongly influence what is measured by the cross-correlations. Knowing that, it is worth checking how different methods measuring cross-correlations depend on autocorrelations in real data.

A straightforward visual inspection of Figure \ref{fig:4} immediately reveals that, on average, most pairs of derivative time series (square symbols) exhibit a weak and constant value of the correlation (as already suggested by the results in Figs. \ref{fig:2} and \ref{fig:3}), while the raw data (circle symbols) shows an increasing trend of cross-correlations for increasing autocorrelations of a given pair.
This tendency can to an extent be reproduced with a VARMA(1,1) model, both simulated and analytical, as presented on the right panel of Figure \ref{fig:4}. The model is defined by two coupled ARMA(1,1) time series with three $2\times 2$ coefficient matrices in total ($A$ for AR---see also Appendix~\ref{app:1}---$B$ for MA and $\Sigma$ for noise).

These observation can be interpreted in several ways. First, correlations can be misleading in the case of raw series, because they might only be a result of autocorrelation. On the other hand, the best fit ARMA(1,1) model, although not fully adequate, cannot explain the whole range of $r_{ij}$ by manipulating $a_1$ when the interactions (i.e., $A_{1,2}$ and $A_{2,1}$ matrix elements) are kept constant. It suggests that $r_{ij}$ of raw series comes at least partly from the functional interactions. Secondly, the $r_{ij}$ of derivatives can be explained more easily by the simple model, which might mean they are less informative. How generic this behavior is, however, remains to be studied analytically. Finally, the $r_{ij}$ of raw BOLD series and of derivatives are uncorrelated, which means that SWPC and MTD methods can produce either complementary or contradictory results that are hard to interpret at the current level of understanding.
More subtle temporal dynamics can be unraveled using the spectral analysis of cross-correlation matrices~\cite{SNAR,LIVAN,TARLAG} and the approach discussed in the next section.

\section{Further testing with surrogate data sets with identical covariance and spectral properties as BOLD}
 \label{sec:bold2}

The properties discussed so far for the raw and derivative time series can be further investigated in the setting of simulated sets that contain the exact same correlation properties as the BOLD data. For that purpose we simulate time series which mimic closely the covariance and spectral properties of the BOLD data. This approach was used recently by Laumann et al.~\cite{Laumann} to study the contribution of various sources of non-stationarity. The starting point is the BOLD data set  from an individual subject.  After applying the bandpass filters usual in most fMRI studies, two estimators are computed: the average power spectrum and the covariance matrix.  After that we create a time series of random Gaussian data of size equal to the real BOLD data set. Subsequently, the spectral content is matched by multiplying these random time series by the two-sided average power spectrum obtained from the real data. Finally, the time series are projected onto the eigenvectors of the covariance matrix calculated from the real data. In this way multivariate data sets---of arbitrary length---are generated, which are stochastic realizations of the chosen BOLD data sets, with identical covariance structure and mean spectral content.

Using this simulation method, similarly as in Sec.~\ref{sec:VII}, we test the scenario of a sudden change in the covariance, considered in the work being discussed \cite{Shine2015}. Figure~\ref{fig:7} illustrates an example using the same BOLD data used in Fig. \ref{fig:4} consisting of 998 time series. Here are plotted the covariances of a pair of time series, both for the raw and the differentiated ones. Up to the step indicated by the dashed line, the series corresponds to a realization of the selected original data (same spectra and covariance) having high cross-covariance, approx. 0.8, and further on to a sudden change of non-diagonal entries of the covariance matrix to 0.2 we introduced in the simulation. In a similar manner, we also introduced simultaneous change in both cross- (again to 0.2) and auto-covariance (to 0.4).
It can be seen that estimators of both raw series and derivatives track the change. Note the relatively smaller amplitude of covariance of derivatives, something expected by definition, as follows from discussion in Sec.~\ref{sec:examples}.
Next, the spread of MTD metric is much greater than that of SWPC, as illustrated by the shaded strips of standard deviations.
At the same time, the intuitions gained from autoregressive processes, as demonstrated in Fig. \ref{fig:6}, appear to hold. The autocovariance does affect the two metrics differently: MTD remains relatively insensitive to it, while it does make the step in SWPC smaller.
It should be emphasized that such simultaneous amplitude modulation does not necessarily arise from any communication of the two ROIs.
Since neither method is able to tell the difference, any preference in such a case is debatable.
 
\begin{figure}[b]
\centering
\includegraphics[width=0.65\textwidth]{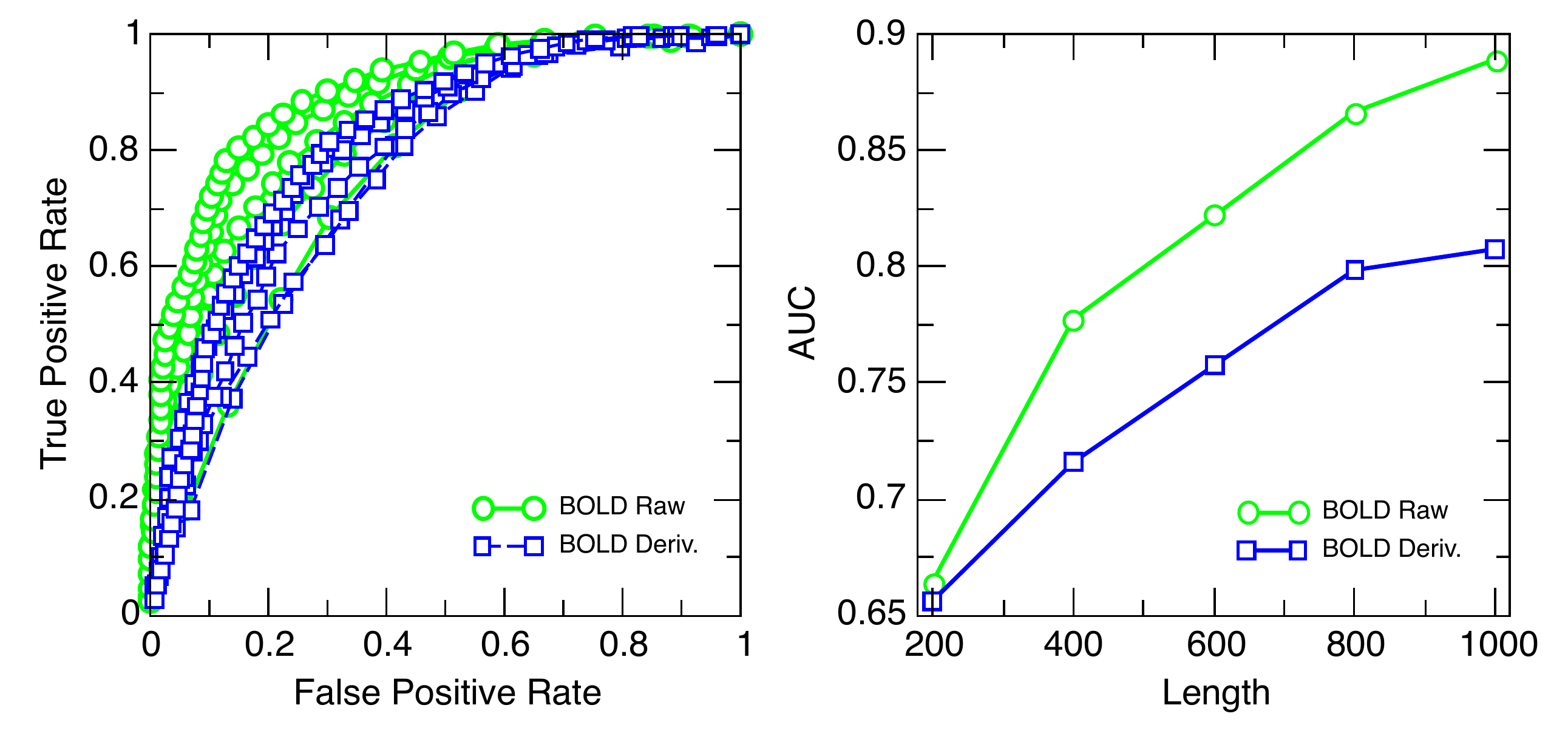}
\caption{\label{fig:5}Left Panel: Receiving Operating Characteristic curves obtained for both methods in detecting the presence of links ($1.75 \sigma$ threshold) computed for the five time series length (indicated in the right panel). Right Panel: The results correspond to the area under the ROC curve, computed for both methods as a function of increasing length of the time series considered. Note that the similarity of performance of the two measures for relatively small windows lengths is expected since both must converge to the 0.5 chance value.}
\end{figure}

\section{Detection of ground truth network from realistic BOLD simulations}
\label{sec:network}
An important claim in~\cite{Shine2015} is the apparent benefit of using the MTD approach to estimate the stationary connectivity structure of a functional network. They compared the performance of the correlation of derivatives against existing methods studying a well characterized fMRI dataset as well as a previously published gold-standard simulated data set~\cite{Smith} obtained from FMRIB (http://www.fmrib.ox.ac.uk/ analysis/netsim). 
For the sake of comparison with the original result, we use the exact same setting, irrespective of its suitability to test for a time-varying connectivity.
The data set is simulated BOLD signals with known network structure, thus enabling evaluation of a range of connectivity models and comparing the results with the ``ground truth'' network structure.
Briefly, this dataset consists of 28 simulations of BOLD data in 50 realizations (TR = 1.5--3 s; 200--1000 individual time points; 5--50 nodes; and differing levels of noise and hemodynamic response function variability). Each simulated dataset was created using an fMRI forward model based on dynamic causal modeling (DCM~\cite{friston}), combined with a non-linear balloon model~\cite{buxton} to simulate vascular dynamics (see~\cite{Smith} for details of the simulation). 

Here we restrict our attention to the network labeled ``sim4'',  which simulates 50 BOLD signals. The simulated network comprises 10 regular modules interconnected by a few links, forming a typical small-word graph. There is a total of 50 nodes interconnected by 61 positive off-diagonal interactions,  40 of them corresponding to nearest neighbors (link weights  are $0.4 \pm 0.03$; mean $\pm$ s.d.). There are also 50 negative (diagonal) self- interactions which determine the characteristic time scale for the dynamics. The data set contains 50 stochastic realizations which meant to represent fMRI records of different human individuals.

We proceed to use the same dataset and check how well the methods perform in predicting from the time series the underlying graph. Specifically, we check how well the correlation matrices obtained from both the raw BOLD dataset and its derivative describe the ground-truth network. We used partial correlations (instead of Pearson's) since that was the optimal method reported in the original Smith et al.~\cite{Smith} report.

To gauge each method we used the receiver operating characteristic curve (ROC~\cite{ROC}), which benchmark specificity and sensitivity as a function of a given parameter. To determine whether a connection between two nodes is predicted or not we choose a decision threshold of 1.75 $\sigma$ at the entries $(i,j)$ of the partial correlation matrices, interpreting any value larger than that as a connection between such nodes. The presence of each link predicted in this way is  compared with the respective entry in the adjacency matrix of the sim4 network, resulting in a false positive or a true positive event. The same procedure is applied for a range of time series lengths and repeated for the raw time series (green points) and the derivatives ( blue points). 

Figure \ref{fig:5} illustrates the results: there is a family of curves corresponding to various lengths of the considered time series. The shortest data (T=200 samples) give the less confident results and the longest (T=1000 samples) a very good estimation of the network connections. The area under the curve (AUC), plotted in the right  panel of Fig.~\ref{fig:5}, is a good estimate of the performance of the method, where a value of 1 corresponds to a perfect prediction and a value of 0.5 is equivalent to chance.
The results in this figure clearly show that the estimates based on the derivatives perform worse than the ones using the raw BOLD time series, suggesting that MTD offers no advantage over standard methods.

\section{Conclusions}
In this paper we have analyzed the basis of a metric dubbed Multiplicative Temporal Derivatives recently proposed as a novel way to determine changes in functional correlations between regions of interest in the brain.
Even though we focused on properties of only two DFC methods, MTD together with sliding-window Pearson correlation,
many others are available and it is noteworthy that there are efforts~\cite{Thompson2017b} to systematically assess their performance.

A formal comparison of the MTD formula with the Pearson correlation of a time derivative reveals two differences, namely that derivatives are not centered in MTD and that their variance is computed over different ensembles. We find it negligible in practice, although centering and windowed standardization tend to decrease uncertainty of estimating correlations.
In effect, what we compared was the mathematical features of correlations of raw and derivative time series. Consequently, the choice of the examined scenarios was dictated by relation of these features to characteristics of BOLD signals.
The results of our analysis show that in a realistic scenario of stationary network detection a metric based on derivatives performs worse.
This could be a consequence of decreased signal-to-noise ratio of such time series---which we expect from the increased variance of noise and decreased amplitude of oscillatory signals for derivatives of a discrete time series---as well as its enhanced stationarity. Derivatives, on the other hand, are not affected by low frequency drifts. 
A comparison with the SWPC of the raw time series in non-stationary set-ups also does not indicate any evident mathematical advantages of the proposed metric over commonly used correlation methods: it reveals lower sensitivity of derivatives to autocorrelations, which might offer higher reliability but at the cost of slower detection and larger deviations. 
The extent and complementarity of information carried by correlations of raw BOLD series and its derivatives demands further inquiry.

\newpage
\begin{acknowledgements}
Work conducted under the auspice of the Jagiellonian University-UNSAM Cooperation Agreement. JKO was supported by the Grant DEC-2015/17/D/ST2/03492 of the National Science Centre (Poland). JKO thanks Valeria Pattacini from the Office of International Relations (ORI) of Universidad de San Mart\'{i}n, UNSAM, (Argentina) for facilitating his visit and the UNSAM's hospitality. WT appreciates the financial support from the Polish Ministry of Science and Higher Education through the "Diamond Grant"  0225/DIA/2015/44.  Authors thank Steve M. Smith and Mark Woolrich (Imperial College, London) for sharing information on the Netsim package. DRC was supported in part by CONICET (Argentina) and Escuela de Ciencia y Tecnolog\'{i}a, UNSAM.  The authors are grateful to Ignacio Cifre for valuable discussions.
\end{acknowledgements}

\appendix

\section{VAR(1)}
\label{app:1}

Let us consider a vector autoregressive process, VAR(q), governed by the following equation
\begin{equation}
x_{it}=\sum_{k=1}^{q}\sum_{j=1}^{N} A^{(k)}_{ij}x_{j,t-k}+\xi_{it}. \label{eq:VAR1}
\end{equation}
The equations \eqref{eq:2series} are a special case of the above process, which can be seen after inserting $q=1$, $N=2$, $A_{11}^{(1)}=A_{22}^{(1)}=a_1$, $A_{12}^{(1)}=A_{12}^{(1)}=a_2$.
In general, there are $N$ time series indexed by $i, j$, and memory of $q$ time steps. The matrices $\bm{A^{(k)}}$ are of size $N\times N$ (in general non-symmetric) with elements $A^{(k)}_{ij}$, and $\xi_{it}$ are Gaussian random variables which are uncorrelated in time, i.e., $\xi_{it}$ and $\xi_{jt'}$ are independent if $t\neq t'$.
General VAR models allow for correlated noise with true covariances ${\rm E} [\xi_{it}\xi_{jt}]=\Sigma_{ij}$, where $\Sigma$ is symmetric positive definite.
By $\< \>$ we denote the sample average of the time series. For instance, in this case the estimator of the true covariance matrix reads $\<\xi_{it}\xi_{jt}\>=\frac{1}{T}\sum_{t=1}^{T}\xi_{it}\xi_{jt}$, where $T$ is the length of the series. For infinite time series length  the true covariance matrix and its estimator  are identical. 

From this point on our aim is to calculate the Pearson correlation coefficients for VAR(1) process and its temporal derivative.
We define the time-lagged cross-covariance matrix as
\begin{equation}
C(\tau)_{ij}:=\< x_{it}x_{j,t+\tau}\>=C(-\tau)_{ji}.
\end{equation}
For brevity we denote $\bm{C}\equiv	\bm{C}(0)$ and $\bm{A}\equiv\bm{A^{(1)}}$. For simplicity of calculation let us start by deriving $C(1)$.
\begin{equation}
C(1)_{ij}=\< x_{it}x_{j,t+1}\>=\<x_{it}\left(\sum_{k=1}^{N}A_{jk}x_{kt}+\xi_{j,t+1}\right)\>=\sum_{k=1}^{N}A_{jk} \<x_{it}x_{kt}\>+\<x_{it}\xi_{jt+1}\>.
\end{equation}
The first equality is the definition, in the second one we substitute \eqref{eq:VAR1} for $x_{j,t+1}$, in the third we expand the multiplication.
The last term vanishes because the only case in which it could give a contribution is when $x_{it}$ contained $\xi_{j,t+1}$---but according to the definition \eqref{eq:VAR1}, it does not.
Moreover, we recognize the first term as the equal-time covariance matrix. Therefore
\begin{equation}
C(1)_{ij}=\sum_{k=1}^{N}C_{ik}A_{jk}. \label{eq:unitlag}
\end{equation}
The results can be presented in a clearer fashion with the use of matrix notation
\begin{equation}
\bm{C}(1)=\bm{C}\bm{A}^{T}, 
\label{VAR1rule}
\end{equation}
where the sum over matrix elements has been replaced with matrix multiplication, and the reversed order of indices of matrix elements ($A_{jk}$ instead of $A_{kj}$) has been replaced with the transpose of the matrix $\bm{A}$ denoted by the superscript $T$.
Such notation provides a considerably more manageable equations.

It is worth noting that the assumption $\<x_{it}\xi_{j,t+1}\>=0$ is true only for long time series $T\to \infty$. The empirical sample average might yield a non-zero result. This might be partly responsible for the discrepancy between analytical values and simulations observed in Fig.~\ref{fig:2}.


The equal-time covariance matrix can be obtained by repeating the above steps:
\begin{equation}
C_{ij}=\<x_{it}x_{jt}\>=\sum_{k=1}^{N}\sum_{l=1}^{N}A_{ik}A_{jl} \<x_{k,t-1}x_{l,t-1}\>+\<\xi_{it}\xi_{j,t}\> = \nonumber \\
\sum_{k=1}^{N}\sum_{l=1}^{N} A_{ik}A_{jl} C_{kl}+\Sigma_{ij},
\end{equation} 
where we already omitted the substitution of \eqref{eq:VAR1}, and 
by virtue of stationarity of the time series $\<x_{it}x_{jt}\>=\<x_{i,t-1}x_{j,t-1}\>$. In matrix notation
\begin{equation}
\label{eq:covar}
\bm{C}=\bm{ACA}^{T}+\bm{\Sigma}.
\end{equation}
We solve this equation in App.~\ref{sec:vectorisation}.

\section{Time derivative of VAR(1)}
\label{app:2}

Let us rewrite \eqref{eq:VAR1} for $q=1$ in matrix notation
\begin{equation}
\bm{x}_{t}= \bm{A} \bm{x}_{t-1}+\bm{\xi}_{t}, \label{eq:dVAR}
\end{equation}
where $\bm{x}_{t}=[x_{1t},\ldots,x_{Nt}]$ and $\bm{\xi}_{t}=[\xi_{1t},\ldots,\xi_{Nt}]$ are $N$-element vectors. By temporal derivative we mean here taking finite forward differences 
\begin{equation}
d\bm{x}_{t}=\bm{x}_{t+1}-\bm{x}_{t}=\bm{A} (\bm{x}_{t}-\bm{x}_{t-1})+\bm{\xi}_{t+1}-\bm{\xi}_{t}.
\end{equation}
To obtain its equal-time cross covariance matrix---which we denote by $\bm{C}'$---we follow the same steps as outlined in the previous section, which leads to
\begin{align}
\bm{C}'&=2\bm{C}-\bm{C}(1)-\bm{C}(-1)\\
&=\bm{C}(\bm{1}-\bm{A}^{\rm{T}})+(\bm{1}-\bm{A})\bm{C},
\end{align}
where the first equality is general, while the second one holds for VAR(1) process only (cf.~\ref{VAR1rule}); $\bm{1}$ denotes the $N\times N$ unit matrix. Thanks to the connection between covariances of the derivatives and the raw time series, now we only need to solve \eqref{eq:covar}, which is described in the next section.
Although more convoluted, equations for $\bm{C}$ and $\bm{C'}$ can be obtained also for higher VAR orders, as well as for VARMA models.

\section{Vectorization}
\label{sec:vectorisation}
Equation \eqref{eq:covar} provides us with a way to compute $\bm{C}$ (in the limit $T \to \infty$) knowing $\bm{A}$ and $\bm{\Sigma}$.
The procedure is simpler if we utilize vectorization (see, e.g.,~\cite{Laub}), which means stacking columns of a matrix into a single column vector. In the vec notation the equation takes the form
\begin{align}
vec(\bm{C})&=vec (\bm{ACA}^T)+vec(\bm{\Sigma})\\
&=(\bm{A}\otimes\bm{A}) vec(\bm{C})+vec(\bm{\Sigma}),
\end{align}
where in the second line  we used the property  $vec(\bm{A_1}\bm{A_2}\bm{A_3})=(\bm{A_3}^T\otimes \bm{A_1}) vec (\bm{A_2})$, which expresses the  matrix multiplication as a linear transformation on matrices with the help of 
 the Kronecker product $\otimes$. Solving for $vec(\bm{C})$ is now straightforward:
\begin{equation}
vec(\bm{C})=\left(\bm{1}\otimes \bm{1}-\bm{A}\otimes \bm{A}\right)^{-1} vec(\bm{\Sigma}).
\end{equation}
When one of the eigenvalues of $\bm{A}$ is equal to one, the matrix in the parenthesis is not invertible (which may be an onset of non-stationarity of time series).

\bibliographystyle{plain}

\end{document}